\documentclass[letterpaper]{jpconf}
\usepackage{graphicx}
\begin{document}
\title{Future Hadron Physics Facilities at Fermilab}

\author{Jeffrey A. Appel}

\address{Fermilab\\PO Box 500, Batavia, IL 60510, USA} 

\ead{appel@fnal.gov}

\begin{abstract}
Fermilab's hadron physics research continues in all its accelerator-based
programs.  These efforts will be identified, and the optimization of the
Fermilab schedules for physics will be described.  In addition to the
immediate plans, the Fermilab Long Range Plan will be cited, and the
status and potential role of a new proton source, the Proton Driver, will
be described.
\end{abstract}.

\section{Hadron Physics at Fermilab}

Hadron physics is found in three areas of accelerator physics at 
Fermilab: the Tevatron collider experiments, the neutrino program, 
and fixed target physics with beams from the Main Injector.  At the 
collider, CDF and DZero are in the midst of a multi-year run now, and BTeV 
is scheduled to take over as soon as these programs are ended.  In the
neutrino program, we have MiniBooNE using beam from the Booster 
accelerator now, and will be turning on NuMI, the neutrino program using 
beam from the Main Injector, in a few weeks.  The MINOS experiment will 
operate with the neutrino beams as soon as they are available, and the 
MINERvA experiment has been approved to join as soon as it can be built.  
Finally, there is the MIPP experiment, E907 - Main Injector Particle 
Production, finishing its commissioning run now, and about to get its 
physics data.  Also approved for the 120 GeV Main Injector program is a 
Drell-Yan experiment (E906), which is starting to build its detector this 
year.  The possibility of returning to run for kaon physics using beam 
from the Main Injector is retained.

The Fermilab accelerator experiment schedules through 2009 are available 
on the web at

\begin{center}
\verb|http://www.fnal.gov/directorate/program_planning/schedule/index.html|
\end{center}

Optimization of running conditions is a challenge with such a diverse 
program, especially since each component of the Fermilab program has its 
own preference for how we should operate the accelerator complex, 
particularly the proton source.

The Tevatron Collider benefits from as rapid a cycling of the upstream 
accelerators as is consistent with antiproton beam-cooling times in the 
Accumulator and Recycler rings, with as much beam as acceptable for 
losses.  The neutrino program can use even more rapid cycle times, with as 
much beam as acceptable for losses.  For neutrinos, there is no lower 
limit on how fast the proton source accelerators should cycle.  At the 
opposite extreme is the so-called SY120/Fixed Target program, which 
depends on experiment and test-beam user status.  During set up, multiple, 
somewhat short duration spills per minute is optimal.  During data taking, 
long cycles with long spills are best for maximizing integrated spill 
seconds by minimizing end effects.  The Office of Program Planning is 
charged with organizing all this by coordinating between experiments and 
the accelerator, under given guidelines from Director.  At the moment, for 
example, there is the following guideline on the use of beam time for 
SY120:

\begin{verbatim}
  The integrated effect of running SY120 beam will not reduce the antiproton 
  stacking rate by more than 5% globally, with the details of scheduling 
  to be worked out between the experimenters and the Office of Program 
  Planning.
\end{verbatim}
Note that this guideline is flexible, leaving the details of spill length
and rate open to maximize the physics of the SY120 program.  We also have 
the flexibility of mixing the various spill configurations in the 
accelerator time line sequence, adding an element of flexibility - and 
complexity.

\section{Fermilab's Long Range Plan}

The Fermilab Director established the Fermilab Long Range Planning 
Committee (LRPC) in the spring of 2003. An excerpt from the charge to the 
LRPC states:

\begin{verbatim}
  I would like the Long-range Planning Committee to develop in detail a few
  realistically achievable options for the Fermilab program in the next decade 
  under each possible outcome for the Linear Collider.
\end{verbatim}
  
It was clear from the start that a new intense proton source to serve 
long baseline neutrino experiments and to provide other new physics 
options at Fermilab was one such option.  A LRPC working group, with Bob 
Kephart as chairman, was charged to explore this option.  The full LRPC 
received recommendations from this working group.  The recommendations 
were subsequently adopted in the final FLRPC report and by the Director.

\section{Fermilab Proton Source}

\subsection{Introduction}

The Fermilab Linear Accelerator and Booster are old machines.  At the same 
time, there is a desire for significantly more intense proton sources for 
long baseline neutrino physics.  In order to meet the new demands, a new 
proton source will need to be constructed.  The high level parameters 
desired for such a new facility at Fermilab are:

\begin{itemize}
\item $0.5-2.0$ $MW$ beam power at 8 Gev
\item $2.0$ $MW$ beam power at 120 GeV, 6 times the power of the current 
Main Injector
\end{itemize}

There are two possible implementations being considered at Fermilab; an 8 
GeV synchrotron and an 8 GeV superconducting rf linear accelerator.  A web 
page has been established to track the progress on such a new proton 
source, the Proton Driver:

\begin{center}
         http://www-bd.fnal.gov/pdriver/
\end{center}

Currently, most of the accelerator effort is focused on bringing design 
for the linear accelerator to the same level as that for the synchrotron 
so that a final choice and proposal can be made.  However, it is safe to 
say that there is particular appreciation for some linear accelerator 
advantages: better performance, flexibility, and the connection and 
possible synergies with an International Linear Collider (ILC), now that 
"cold" technology has been chosen for the latter.

\subsection{Main Injector Upgrades}

For either choice of 8 GeV injector (synchrotron or linear accelerator) 
the beam in the Main Injector will increase by a factor of ~5 from its 
design value of $3.0$ $10^{13}$ protons per pulse to $\sim1.5$ $10^{14}$ 
protons per cycle.  The Main Injector beam power could also be increased 
by shortening the Main Injector ramp time.  This would require additional 
magnet power supplies, and could be done prior to a Proton Driver as a 
first step.  Both more protons per cycle and faster ramp times imply more 
rf power, meaning money is required.

\subsection{Proton Driver Status and Plans}

The recent International Technology Recommendation Panel opted for a 
"cold" technology for the International Linear Collider. This 
recommendation, and its wide acceptance, has provided a huge boost for a 
super-conducting rf linear-accelerator-based Proton Driver at Fermilab.  
Fermilab has been working on both warm and cold linear-collider-related 
technology.  In FY 2005, all this effort will be dedicated to cold 
technology, also advancing a linear-accelerator Proton Driver.  So, FY 
2005 will see as much as a factor of two increase in super-conducting R\&D 
spending at Fermilab relative to FY 2004.  Plans are forming for a SCRF 
Module Test Facility (SMTF) to be built in the Meson East area at 
Fermilab.  Long lead-time items like modulators are already being ordered.  

\subsection{Time Scale for a Proton Driver}

It seems likely that a new, more intense proton source will be proposed 
for construction at Fermilab in the relatively near future.  Such a
project would be similar in scope to the Main Injector Project (i.e., for
cost and schedule).  An 8 GeV synchrotron and a superconducting rf linear
accelerator both appear to be technically possible.  The linear
accelerator is strongly preferred by many, if it can be made affordable.
The Fermilab management has requested that an 8 GeV linear accelerator
design be developed, including cost and schedule information.  Fermilab's
Bill Foster is leading a team to develop the technical design.

It is always hard to guess about time scales.  A technically limited 
schedule (when was the last time this happened?) could have the following 
DOE construction project critical decision (CD) schedule:

\begin{itemize}
\item CD-0 (statement of need) submitted in 2005 
\item CD-1 (preliminary acquisition strategy, project execution plan, 
conceptual design report, project scope, baseline cost/schedule range, 
project management plan, hazard analysis, etc.) submitted in 2006
\item CD-2/3a (project baseline approved, approval to start construction) 
in 2007-8.
\end{itemize}

As usual, the availability of funding from DOE may push this schedule 
later.  Once funding is approved, typical projects of this scale (the 
Fermilab Main Injector, SLAC B-factory, Spallation Neutron Source) have 
construction times of 4-5 years.  The time scale will also depend on how 
the International Linear Collider plays out over the next few years (e.g., 
could the Proton Driver be seen as an ILC Engineering Test Facility?).  

\section{Proton Driver - The Physics Case}

\subsection{Introduction}
It is up to users to make the physics case that a Proton Driver is 
required, and that it should be built as soon as possible.  Making the 
physics case is crucial in all of this!  A workshop to think about the 
broadest physics potential of a Proton Driver was just held.  The program, 
with links to talks can be found at the above Proton Driver web page.

Steve Geer has been asked to organize the documentation of the physics 
case.  A strong physics case, coupled to a machine technical design, will 
make it possible to submit a Proton Driver project to the DOE for approval 
and funding.  The recent Proton Driver Workshop was a step in this 
process.  Information on the workshop and talks can be found linked to the 
web page:

\begin{center}
     http://www-td.fnal.gov/projects/PD/PhysicsIncludes/Workshop/index.html
\end{center}

Three of the working groups at the workshop may be most interesting for 
hadron physics:  

\begin{itemize}

\item WG2 on Neutrino Interactions -- Jorge Morfin (Fermilab), Rex Tayloe 
(Indiana), Ron Ransome (Rutgers), conveners
\item WG4 on Kaons/Pions --
Hogan Nguyen (Fermilab) and Taku Yamanaka (Osaka), conveners
\item WG5 on Antiprotons -- Dave Christian (Fermilab) and Mark Mandelkern 
(UC-Irvine), conveners
\end{itemize}
While not represented at the workshop, there is also a working group on 
neutrons.  The workshop, of course, had additional working groups less 
connected to hadron physics.

In each case, the working groups identified physics measurements of 
interest to the particle and nuclear physics communities in the era of a 
Proton Driver.  However, in principle, some of the measurements discussed 
at the workshop could be made sooner.  Organizing to make these earlier
measurements may focus a part of the research community in anticipation of 
a Proton Driver.  It would be easier then to do the experiments which 
depend on the flux available from a Proton Driver.

\subsection{Neutrino Working Group}

The Neutrino Working Group included the following physics topics in its 
discussion of goals:
 
\begin{itemize}         
\item Strange quark contribution and axial form factor through parity 
violation
\item Duality
\item Coherent pion production
\item Exclusive resonant final states
\item Parton distribution functions, particularly at high $x_Bj$
\begin{itemize}
\item Generalized parton distribution functions including quark flavor 
        dependence through the weak analog of deeply-virtual Compton 
        scattering $n(\nu,\gamma)p$
\item Detailed study of nuclear effects with neutrinos that are expected 
        to be significantly different than those measured with charged 
        leptons
\end{itemize}
\item Spin-dependent distribution measurements using polarized targets
\item A-dependence and more detailed nuclear effects
\end{itemize}

\subsection{Kaon and Pion Working Group}
In the case of the Kaon and Pion Working Group, the following were among 
the topics discussed:

\begin{itemize}
\item Production dynamics for exotic/rare quark and gluon states
\item Rare pion decays, e.g., improving $V_{ud}$ from pion beta decay in 
flight
\item CKM parameters via $K \to \pi \nu \overline{\nu}$
\item Pushing further in $K \to \pi l l$ in the search for new physics
\end{itemize}
Many of these measurements can be expected to help define the nature of 
the high mass objects which are anticipated to be found at the Tevatron or 
the LHC.

\subsection{Proton Driver Study - Next Steps}

The Proton Driver Working Group draft reports will go to a Proton Driver 
Science Advisory Committee (PDSAC)\cite{PDSAC}, whose chair is Peter 
Meyers (Princeton).  The PDSAC is composed of particle and nuclear 
physicists, and is international in composition.  The plan is to have the 
PDSAC review conveners' presentations and the reports in order to advise 
the Laboratory on the strongest physics motivations for a Proton Driver.  
A draft statement of the physics case (a statement of need) could be ready 
for reading over the winter holidays.  Then, a final draft statement of 
the physics case could be prepared for submission to DOE.

\section{Final Comments}

I have focused on a only part of the hadron physics in Fermilab's future.  
This part is dominantly proposal driven.  I have not said much about the 
collider or MIPP, which are covered in separate talks at this First 
Meeting of the APS Topical Group on Hadron Physics.  In its vision of the 
future at Fermilab, the Long Range Plan says

\begin{verbatim}
  Fermilab will remain the primary site for accelerator-based particle 
  physics in the U.S. in the next decade and beyond. 
\end{verbatim}
Defining the future of much of hadron physics at Fermilab, e.g., use 
of neutrinos, hadron beams, and even Tevatron beams will depend on the 
users.  What will be the total available beam?  Will we build a Proton 
Driver?  What fraction of the available protons will go to hadron physics?  
Will a stretcher ring be proposed and built?  

A robust program depends on robust interest, and eventually, compelling 
physics-driven experiment proposals.  We may hope this meeting of the 
Topical Group on Hadron Physics, and related gatherings in the future, 
will contribute directly to that robust program.

\ack

Much of the material on the Proton Driver is taken from the presentation 
by Bob Kephart on October 6, 2004, to the Proton Driver workshop at 
Fermilab.  Thanks also go to David Christian, Hugh Montgomery, Hogan 
Nguyen, and Jorge Morfin for comments on a draft of this paper.


\section*{References}
\begin{thebibliography}{1}
\bibitem{PDSAC} The membership of the Proton Driver Science Advisory 
Committee is:
\begin{itemize}
\item Peter Meyers, Princeton University, USA, chair
\item Ed Blucher, University of Chicago, USA
\item Gerhard Buchalla, Ludwig-Maximilians University, Munich, Germany
\item John Dainton, University of Liverpool, UK 
\item Yves Declais, LAPP-IN2P3-CNRS Annecy, France
\item Lance Dixon, SLAC, USA
\item Umberto Dosselli, INFN and University of Padua, Italy
\item Don Geesaman, Argonne National Laboratory, USA
\item Geoff Greene, Oak Ridge National Laboratory, USA
\item Taka Kondo, KEK, Japan
\item Marvin Marshak, University of Minnesota, USA
\item Bill Molzon, University of California-Irvine, USA
\item Hitoshi Murayama, University of California-Berkeley, USA
\item Jim Siegrist, University of California-Berkeley, USA
\item Tony Thomas, Thomas Jefferson National Laboratory, USA
\item Taku Yamanaka, Osaka University, Japan
\end{itemize}
\end{thebibliography}
\end{document}